\def\smalllinestretch{\renewcommand{\baselinestretch}{0.7}}
\def\largelinestretch{\renewcommand{\baselinestretch}{1.0}}
 \def\tr {\, \mbox{tr} \,}
 \def\Trp{\, \mbox{Tr}^{\prime} \,}
 \def\trp{\, \mbox{tr}^{\prime} \,}

 \def\L{{\cal L}}

 \def\al{{\alpha}}
 \def\be{{\beta}}

 \def\Fi{\Phi}

 \def\gev{\,\mbox{GeV}}

\def\square{\hbox{\vrule\vbox{\hrule\phantom{o}\hrule}\vrule}}

\documentstyle{article}

\def\largelinestretch{\renewcommand{\baselinestretch}{1.2}}
\largelinestretch\small\normalsize
\title{
        Heat-kernel calculation of quark determinant \\
                and computer algebra
      }
 \author{
A.A.Bel'kov${}^1$,
D.Ebert${}^2$,
A.V.Lanyov${}^1$,
A.Schaale${}^3$
\\
\\
\small
${}^1$
        Particle Physics Laboratory, Joint Institute for Nuclear Research,
\hfill\\
\small
        Head Post Office, P.O.Box 79, 101000 Moscow, Russia
\hfill\\
\small
${}^2$
        Institut f\"ur Elementarteilchenphysik, Humboldt-Universit\"at,
\hfill\\
\small
        Invalidenstra\ss e 110, O-1040 Berlin, Germany
\hfill\\
\small
${}^3$
        DESY-Zeuthen
        Platanenallee 6, O-1615 Zeuthen, Germany
\hfill\\
}
\begin{document}
\largelinestretch\normalsize
   \thispagestyle{empty}
   \begin{titlepage}
   \thispagestyle{empty}
   \maketitle
   \begin{abstract}

   In this paper there we describe the calculational background of
deriving a strong  meson Lagrangian from the
Nambu--Jona-Lasinio quark model using the computer algebra
systems FORM and REDUCE in recursive algorithms, based on the heat-kernel
method for the calculation of the quark determinant.
   \end{abstract}
   \end{titlepage}

   Computer Algebra Systems (CAS) such as FORM \cite{form}  or
REDUCE \cite{reduce}  have been successfully used in the high
energy physics for a long time, especially in the  field  of
the perturbative theory calculations  of  higher  order contributions
in the standard model.
  Besides  these  ``classical'' fields of application of CAS, they
can  be  also  used  for  solving various problems connected
with  non-renormalizable  models  such as chiral meson models.
   In the following we describe the calculational background
of the recent derivation \cite{91-08} of meson Lagrangian
describing strong interactions of scalar, pseudoscalar, vector
and axial-vector fields.

   The strong interactions of quarks at high energies
($E \gg 1 \gev$) are  described  by  the  Quantum  Chromodynamics
(QCD) \cite{qcd}.  The QCD Lagrangian has following form:
$${\cal L}_{\hbox{\scriptsize{QCD}}} =
\sum_{f} \bar{q}^{f}(i\widehat{D} - m^{0}_{f})q^{f}
           - {1\over 4} \, G^{\mu\nu} \, G_{\mu\nu}$$
with
$$
G_{\mu\nu} = \partial_\mu G_\nu -\partial_\nu G_\mu  + i g[G_\mu ,G_\nu ],
$$
$$\widehat{D} = \gamma^\mu (\partial_\mu  - igG_\mu  ) ,\quad
       G_\mu  = \sum^{}_{a} G^{a}_\mu  {\lambda^{a}_{C}\over 2},$$
where $q$ is the quark field, $G_\mu $ is the gluon field,
$g$  is  the  coupling constant,
$m^{0}_{f}$ is the current  quark  mass,
$\gamma^\mu $  are  the  Dirac matrices,
$\lambda_{C}$ the $SU(3)_C$ matrices.

Some special properties of the QCD like  the  confinement  of
quarks and the momentum dependence of the strong coupling constant
$\alpha_{s}(q^{2})$ which becomes large at low momenta
($\alpha_{s} \approx 1$) make it impossible  to use the QCD
for an adequate description of meson processes at  low energies.

Until now it was not possible to derive a meson  theory  from
the QCD. On the other side, meson processes have been studied in
a framework of chiral Lagrangians since 1960s. One possible  way
to built up a QCD-based chiral meson theory is  to  start  with
some effective  quark  Lagrangian, e.g.   with   the
Nambu--Jona-Lasinio model (NJL) \cite{njl}.

The NJL Lagrangian of the  effective  four-quark  interaction
has the form:
\begin{equation}
{\cal L}_{\hbox{\scriptsize{NJL}}} = \bar{q}(i\widehat{\partial }-m_{0})q
+ {\cal L}_{int}
\label{njl-lagr}
\end{equation}
with
$$
{\cal L}_{int} = 2G_{1}\left\{(\bar{q}{1\over 2}\lambda^{i}_{f}q)^{2}
  +(\bar{q}i\gamma^{5}{1\over 2}\lambda^{i}_{f}q)^{2}\right\}
-2G_{2}\left\{(\bar{q}\gamma^\mu {1\over 2}\lambda^{i}_{f}q)^{2}
  +(\bar{q}\gamma^\mu \gamma^{5}{1\over 2}\lambda^{i}_{f}q)^{2}\right\},$$
$G_{1}$ and $G_{2}$ are some empirical constants.

   Let us show that the  NJL model can be motivated from the
nonperturbative QCD
using some assumption about  the infrared behavior of gluons.
The generating functional of QCD  has the form
\cite{ebert-reinhardt}:
$$ {\cal Z} = \int {\cal D} q {\cal D} \bar{q} {\cal D} G
\exp \left\{i\int d^{4}x
           \left[\bar{q}^{f}(i\widehat{\partial }-m^{0}_{f})q^{f}
               + g\bar{q}^{f}\widehat{G}^{a}{\lambda^{a}_{C}\over 2}q^{f}
               - {1\over 4}G^{\mu\nu}G_{\mu\nu}\right]\right\}$$
After the integration over the gluon fields one will get
$${\cal Z} = \int {\cal D} q {\cal D} \bar{q} \exp
        \left\{i\int d^{4}x\bar{q}(i\widehat{\partial }-m^{0}_{f})q\right\}
          \exp (W[j])$$
with
$$j^{a}_\mu (x) = \bar{q}(x)\gamma_\mu \lambda^{a}_{C}q(x)$$
as flavour-singlet quark current and $W$ becomes
$$W[i] =\sum^{\infty }_{n=2} {1\over n!}\int
      d^{4}x_{1}...d^{4}x_{n}
      D^{a_{1}...a_{n}}_{\mu_{1}...\mu_{n}}(x_{1},...,x_{n})
      \prod^{n}_{i=1}j^{a_{i}}_{\mu_{i}}(x_{i})$$
The functions $D^{a_{1}...a_{n}}_{\mu_{1}...\mu_{n}}$ contain all
information about  the  gluons.
An analytical integration is not  possible  because  the  behavior
of the gluon propagator and the running coupling constant at  long
distances are  unknown.

Neglecting higher order Green functions one can get some local
approximation for the gluon  propagator
$D_{\mu\nu}(x,y) \propto g_{\mu\nu}\delta^{4}(x-y)$
and find some theoretical base for the approximation
$ {\cal L}_{\hbox{\scriptsize{QCD}}} \rightarrow
  {\cal L}_{\hbox{\scriptsize{NJL}}}$.
In this approximation it is impossible to predict  the  values  of
the constants $G_{1}$ and $G_{2}$  from  QCD  principles.  They  are  to  be
determined experimentally.

One can see that the group structure of the QCD
$$ SU(3)_C^{\hbox{\scriptsize local}} \otimes
   SU(N_{f})_{L} \otimes
   SU(N_{f})_{R} \otimes
   U(1) \otimes S $$
and that of the NJL model
$$ SU(3)_C^{\hbox{\scriptsize global}} \otimes
   SU(N_{f})_{L} \otimes
   SU(N_{f})_{R} \otimes
   U (1)\otimes S $$
are very similar ($S$ presents the set of discrete symmetries as $C$,
$P$ and $T$ conjugation).

After this introduction of the NJL model we will now derive a
meson Lagrangian from the NJL Lagrangian (\ref{njl-lagr}) using  path  integral
techniques. First one has to introduce collective fields with  the
form:
$$ S = \sum  S_{i}{1\over 2}\lambda_{i} ,\enspace
   P = \sum  P_{i}{1\over 2}\lambda_{i} ,\enspace
   V^\mu  = -i \sum  V^\mu_{i}{1\over 2}\lambda_{i} ,\enspace
   A^\mu  = -i \sum  A^\mu_{i}{1\over 2}\lambda_{i}
$$
which will correspond  later  to  scalar  (S),  pseudoscalar  (P),
vector (V) and  axial  (A)  colourless  fields.  The  following
notations have been used:
$$ S_{i} = -4G_{1}\bar{q}{1\over 2}\lambda_{i}q ,\enspace
   P_{i} = -4G_{1}\bar{q}i\gamma^{5}{1\over 2}\lambda_{i}q, \enspace
   V^\mu_{i} = -4G_{2}\bar{q}\gamma^\mu {1\over 2}\lambda_{i}q ,\enspace
   A^\mu_{i} = -4G_{2}\bar{q}\gamma^\mu \gamma^{5}{1\over 2}\lambda_{i}q.$$
Now the Lagrangian (\ref{njl-lagr}) can be redefined:
$${\cal L}_{\hbox{\scriptsize{NJL}}} = -{1\over 4G_{1}}\tr\phi^{+}\phi
                            -{1\over 4G_{2}}\tr(V^{2}_\mu +A^{2}_\mu )
   + \bar{q}\left\{i\widehat{\partial}+\widehat{V}+\widehat{A}\gamma^{5}
              -P_{R}(\phi +m_{0})+P_{L}(\phi^{+}+m_{0})\right\}q$$
with $\phi =S+i P$,
$\widehat{V}=V^\mu \gamma_\mu $,
$\widehat{A}=A^\mu \gamma_\mu $
and $P_{R/L}={1\over 2}(1\pm \gamma^{5})$   as   chiral
projectors. This substitution transforms the Lagrangian
(\ref{njl-lagr}) into a bilinear form in quark fields.
   Now  the  integration
over  quark  fields  becomes possible and the generating
functional of the NJL model is

$$Z = \int {\cal D}\phi {\cal D}\phi^{+}{\cal D}V{\cal D}A
\exp \left[i\int d^{4}x - {1\over 4G_{1}}
\tr\phi^{+}\phi - {1\over 4G_{2}}\tr(V^{2}_\mu +A^{2}_\mu ))\right]
\cdot {\cal Z}_{f}$$
with
\begin{equation}
 Z_{f}(\phi ,\phi^{+},V,A) =
  \int {\cal D}q{\cal D}\bar{q} \exp (i\int d^4x \bar{q}i {\bf \widehat{D}}q)
          = (\det i{\bf \widehat{D}}) \label{quark-det}
\end{equation}
and
\begin{eqnarray*}
 i\widehat{\bf D} &=& i \widehat{\partial}+\widehat{V}+\widehat{A}\gamma^5
      -P_{R}(\phi +m_{0}) +P_{L}(\phi^{+}+m_{0})
\\
&=& i (\widehat{\partial} +\widehat{A}_R) - (\phi    +m_0) P_R
   +i (\widehat{\partial} +\widehat{A}_L) - (\phi^{+}+m_0) P_L
\end{eqnarray*}
as Dirac operator. $\widehat{A}_{R/L}=\widehat{V}\pm \widehat{A}$  are
right-  and  left-handed  vector fields.  The  quark
determinant (\ref{quark-det}),   describing   the
interactions of mesons, can be evaluated in quark loops
\cite{kikkawa,volkov-ebert}.

At this place it makes sense to refer to some different
ways in the calculation of the quark-determinant
(\ref{quark-det}). In contrary to the ''straight'' method of
calculating the quark loops \cite{kikkawa} we have used the
heat-kernel method \cite{ebert-reinhardt}. Of course, the
physical results must not depend on the method of calculation
but there are some technical reasons for using this method here.
The main advantage of this method is  that  the
recursive  algorithms  of  this method can be adapted on CAS
quite effectively.  The  infinite contributions can be
conveniently separated too.

The  absolute  value  of  the  quark determinant (\ref{quark-det}) in
``proper-time'' regularization is defined as
\begin{equation}
\log |\det i{\bf \widehat{D}}| \,  =
 - {1\over 2} \Trp \log({\bf \widehat{D}}^{+}{\bf \widehat{D}}) =
 - {1\over 2} \int^{\infty }_{1/\Lambda^{2}} d\tau{1\over \tau}
                    \Trp \exp (-{\bf \widehat{D}}^{+}{\bf \widehat{D}\tau})
\label{logarithm}
\end{equation}
with $\Lambda $ as regularization parameter. The trace $\Trp$ is
to be understood as a space-time integration and a ``normal''
trace over Dirac, colour and flavour indices:
$$ \Trp = \int d^{4}x \trp, \quad
   \trp = \mbox{tr}_C \cdot
          \mbox{tr}_f \cdot
          \mbox{tr}_\gamma.$$
The operator ${\bf \widehat{D}}^{+}{\bf \widehat{D}}$ can be presented as
$${\bf \widehat{D}}^{+}{\bf \widehat{D}} = d_\mu d^\mu  + a(x) +
\mu^{2}\qquad$$
with
$$ d_\mu =\partial_\mu +\Gamma_\mu, \quad
   \Gamma_\mu =V_\mu +A_\mu \gamma^{5}, \quad
   a(x) = i\widehat{\nabla }H + H^{+}H
              + {1\over 4}[\gamma^\mu ,\gamma^\nu ]\Gamma_{\mu\nu} - \mu^{2}.$$
Here $\mu$ plays the role of some free parameter which  will  fix  the
regularization in the region  of  low  momenta.  Below it  will  be
identified to the constituent quark mass. We  used  the  following
notations:
$$ H=P_{R}(\phi +m_{0})+P_{L}(\phi^{+}+m_{0})=S+i\gamma_{5}P, $$
$$ \Gamma_{\mu\nu}=[d_\mu ,d_\nu ]=\partial_\mu \Gamma_\nu -\partial_\nu
\Gamma_\mu +[\Gamma_\mu ,\Gamma_\nu ]
=F^{V}_{\mu\nu}+\gamma^{5}F^{A}_{\mu\nu} ,$$
with $F^{V,A}_{\mu\nu}$ as field strength tensors
$$ F^{V}_{\mu\nu}=\partial_\mu V_\nu -\partial_\nu V_\mu +[V_\mu ,V_\nu
]+[A_\mu ,A_\nu ],$$
$$ F^{A}_{\mu\nu}=\partial_\mu A_\nu -\partial_\nu A_\mu +[V_\mu ,A_\nu
]+[A_\mu ,V_\nu ]$$
and
$$\nabla_\mu H=\partial_\mu H+[V_\mu ,H]-\gamma^{5}\{A_\mu ,H\}$$
as covariant derivative.

The main idea of the heat-kernel method is to evaluate
$$<x\mid \exp (-{\bf \widehat{D}}^{+}{\bf \widehat{D}\tau})\mid y>$$
around its nonperturbated part
$$<x\mid \exp (-(\hbox{\square }+\mu^{2})\tau)\mid y>
= {1\over (4\pi \tau)^{2}} e^{-\mu^{2}\tau+(x-y)^{2}/(4\tau)}$$
in  powers  of proper-time $\tau$  with  the  so-called   Seeley--deWitt
coefficients $h_{k}(x,y)$
$$<x\mid \exp (-{\bf \widehat{D}}^{+}{\bf \widehat{D}\tau})\mid y>
= {1\over (4\pi \tau)^{2}} e^{-\mu^{2}\tau+(x-y)^{2}/(4\tau)}
\sum^{}_{k} h_{k}(x,y)\cdot \tau^{k}.$$
After integration over $\tau$ in (\ref{logarithm}) one gets the
following  expression for $\log \mid \det i \bf \widehat{D}\mid$
$${1\over 2}\log (\det {\bf \widehat{D}}^{+}{\bf \widehat{D}})
= - {1\over 2} {\mu^{4}\over (4\pi )^{2}}
\sum^{}_{k} {\Gamma (k-2,\mu^{2}/\Lambda^{2})\over \mu^{2k}}\Trp h_{k}$$
with
$$\Gamma (\alpha,x)=\int^{\infty }_{x} d t \, e^{-t}t^{\alpha-1}$$
as incomplete gamma function. Using the definition of  the  gamma
function one can separate the divergent and finite parts:
$$
{1\over 2}\log (\det {\bf \widehat{D}}^{+} {\bf \widehat{D}})
= B_{\hbox{pol}}+ B_{\hbox{log}}+ B_{\hbox{fin}}.$$
Here
$$B_{\mbox{pol}} = {1\over 2} {e^{-x} \over (4\pi )^{2}}
 \left[- {\mu^{4}\over 2x^{2}} \Trp h_{0}
          + {1\over x}({\mu^{4}\over 2} \Trp h_{0}
                       -\mu^{2}\Trp h_{1}
                      )\right]$$
has a pole at $x=\mu^2 / \Lambda^2=0$,
$$B_{\hbox{log}} = - {1\over 2} {1\over (4\pi )^{2}}\Gamma (0,x)
    \left[ {1\over 2}\mu^{4}\Trp h_{0}-\mu^{2}\Trp h_{1}
            +\Trp h_{2} \right].$$
is logarithmic divergent.
The finite part has the form:
$$B_{\hbox{fin}} = - {1\over 2} {1\over (4\pi )^{2}}
                      \sum_{k=2}^{\infty} \mu^{4-2k}\Gamma (k-2,x)\Trp h_{k}.$$
The main technical problem now is to calculate the
Seeley--deWitt coefficients $h_{k}$.
   One can find the calculated
heat coefficients up to $k=3$ \cite{kikkawa} and with
simplifications up to $k=6$  in \cite{ebert-reinhardt}.
   In this paper we have calculated the full coefficients up to the
order $n=4$ and and also present the minimal parts of the
heat-coefficients $h_{5,6}$.
Let us show  the general heat-kernel techniques in detail:
The determinant  of  the  positive  definite  operator  ${\bf A}$  is
defined in proper-time regularization by  the  following  integral
relation
$$\log (\det{\bf A}) = - \int^{\infty }_{1/\Lambda^2} {d\tau\over \tau}
\Trp K(\tau) ,\label{log-det-a}$$
where $K(\tau)=e^{-{\bf A}{\tau}}$ is the so-called ``heat kernel''
which  satisfies  the equation
$${\partial \over \partial \tau}K(\tau) + {\bf A}K(\tau) = 0$$
with the boundary condition
$$K(\tau=0) = 1.$$
In the case  discussed  in  the  paper  the  operator  ${\bf A}$  has  the
structure
$${\bf A} = d_\mu d^\mu  + a(x) + \mu^{2} ,$$
where
$$d_\mu  = \partial_\mu  + \Gamma_\mu $$
contains an differential operator
and $a(x)$ does not.

The  asymptotic  behavior  of  ${\bf A}$   at   long   distances
is corresponding with  the  infinite  part  of $\log (\det A)$,
which  is defined by the ``free'' part
$${\bf A}_{0} =\hbox{\square  }+ \mu^{2} , \quad
\hbox{\square} \equiv  \partial_\mu \partial^\mu.$$
Using the ansatz
$$K = K_{0}H$$
it is convenient to separate the ``free''  part $K_{0}$  from  the  heat
kernel. In coordinate representation $K_{0}$ reads
$$K_{0}(x,y;\tau) \equiv  <x\mid K_{0}\mid y> = {1\over (4\pi \tau)^{2}}
 e^{-\mu^{2}\tau+(x-y)^{2}/(4\tau)}$$
and also satisfies the equation
$${\partial K_{0}\over \partial \tau} + {\bf A}_{0}K_{0} = 0 ,$$
with the boundary condition
$$ K_{0}(\tau=0) = 1.$$
The  ``interaction''  part $H$  of  the  heat  kernel  satisfies  the
equation
$$( {\partial \over \partial\tau} + {1\over\tau} z_\mu d^\mu
+ d^\mu d_\mu + a)
          H(x,y;\tau) = 0 ,$$
\begin{equation}
H(x,y=x;\tau=0) = 1 ,
\label{heat-eq}
\end{equation}
where $z_\mu =x_\mu -y_\mu $ and the differential operator
$d_\mu $ acting only on x.
Using now an expansion for $H(\tau)$ in powers of $\tau$
$$H(x,y;\tau) = \sum_{k=1}^{\infty} h_{k}(x,y)\cdot \tau^{k}\label{h}$$
one will get from (\ref{heat-eq}) the recursive relation
$$( n+1 + z_\mu d^\mu  ) h_{n+1} + ( a + d_\mu d^\mu  ) h_{n} = 0
\label{rec-rel}
$$
with boundary condition
\begin{equation}
z_\mu d^\mu h_{0} = 0 .
\label{bound2}
\end{equation}
The heat  coefficients $h_{k}(x)=h(x,y=x)$  are  defined  by  the
diagonal part of the recursive relation (\ref{heat-eq})
\begin{equation}
n h_{n}(x) = - [d_\mu d^\mu  h_{n-1}(x,y)]\mid_{y=x} - a h_{n-1}(x)
\label{rec-rel-second-one}
\label{h2}
\end{equation}
which gives for $n=0$
$$h_{1}(x,x) = - ( d_\mu d^\mu  + a ) h_{0}(x,x) .$$
Here we use the more convenient notations
$$d_\alpha \ldots
d_\beta h_{n}(x,x) \equiv  [d_\alpha \ldots
d_\beta h_{n}(x,y)]\mid_{y=x} .$$
   The expression (\ref{h2}) can be trivially calculated
once the $(n-1)^{th}$ order heat coefficient is known.
   To find the derivative terms $d_\mu d^\mu h_{n-1}$ one
applies the operator $d_\mu d^\mu$ on the recursion relation
(\ref{h2}) for $n-1$ and puts afterwards $y=x$.
   This introduces derivatives up to the order of $h_{n-2}$,
and in addition derivatives of $a$.
   One continues to apply repeatedly operators $d_\mu$ on the
recursion relation (\ref{h2}) for smaller and smaller n
until the desired $n^{th}$ order heat coefficient is completely
expressed in terms differing only by a permutation of the
$d_\mu$'s.
   To get an explicit expression for the multiple derivatives of
$h_n$ one has to bring the $d_\mu$'s in all of these terms into
the same order.
   The reordering of the $d_\mu$'s introduces multiple
commutators of the form:
\begin{equation}
 K_{\mu\nu} = [d_\mu ,d_\nu ] =
     \Gamma_{\mu\nu} ,\hspace{0.3cm} K_{\lambda \mu\nu} =
     [d_{\lambda },K_{\mu\nu}] ,\hspace{0.3cm} K_{\kappa \lambda \mu\nu} =
     [d_{\kappa },K_{\lambda \mu\nu}]\hspace{0.3cm} \hbox{, etc.}
\label{23}
\end{equation}
   Similarly one has to bring the $a$'s to the left of all $d_\mu$'s.
   This introduces the multiple commutators
\begin{equation}
 S_\mu  = [d_\mu ,a] ,\hspace{0.3cm} S_{\mu\nu} =
        [d_\mu ,S_\nu ] , \hspace{0.3cm}S_{\lambda \mu\nu} =
        [d_{\lambda },S_{\mu\nu}] \hspace{0.3cm} \hbox{, etc.} \label{24}
\end{equation}
   In this way the heat coefficients can be completely expressed
in terms of multiple derivatives of $h_0$, of multiple
commutators (\ref{23}) and (\ref{24}), of lower-order heat
coefficients, and of $a$'s, with the latter ones being always
on the left of the derivative operators $d_\mu$.
   The multiple derivatives
$$d_\alpha \ldots
d_\nu h_{0}(x) \equiv  d_\alpha \ldots
d_\nu h_{0}(x,y)\mid_{y=x}$$
are finally calculated by applying the $d_\mu$'s repeatedly on
the differential equation (\ref{bound2}) defining $h_0$.

   Following the strategy outlined above the calculation of the
heat coefficients is straightforward but cumbersome.
   The very lengthy calculations can be performed only by
computer support.
   The calculation of the heat-coefficients is a recursive
process which can be done by CAS very conveniently.

   For the calculations of the heat coefficients we used the CAS
REDUCE and FORM.
   The reason for the use of different CAS for the solution of
one problem is connected with the special features of the
systems.
   FORM is designed for doing fast arithmetic with large
formulae and a special attention was to implement there
structures commonly used in high energy physics, e.g.\ summation
over dummy indices.
   It easy easy to control even difficult calculation processes step
by step since operations are only done on request.
   However, the scope of FORM operations is limited and if some
particular sphere of symbolic calculations is somewhat biased
from the generally used area of multiloop calculations or Dirac
algebra, then a need for unimplemented operations can arise.
    In our case of calculations with nonrenormalisable chiral
models in low-energy meson physics there is a need not only for
summation over some dummy indices but also for the following
operations.

    First of all one needs the operation of cyclic shifting of non-commuting
operators products under the trace sign.
    This operator is required since the following parts of
Lagrangian are equivalent:
$$
\L_1= \tr \left( \partial_\mu\Fi \cdot \Fi \; \partial_\mu\Fi \cdot \Fi \right)
 \quad \mbox{and} \quad
\L_2= \tr \left( \Fi \; \partial_\mu\Fi \cdot \Fi \; \partial_\mu\Fi \right)
$$
   and one should transform them to some unique form, either the
former or the latter one.

   Secondly, one way of testing the obtained physical results is
to test their symmetry properties.
   In our case we have Hermitian conjugation and the
transformation of intrinsic parity.
   One needs some means to test automatically the symmetry
properties of the results.

   Both FORM and REDUCE lack the described operations.
   FORM has no possibility of adding new non-trivial operations.
   For this purpose we used CAS REDUCE which is a completely
open system allowing the user to access directly the internal
data structure and implement new operations and data types by
the ``symbolic'' style programming.
   We implemented the required operations absent in FORM and
also such as summation over dummy indices.
   In such a way the whole work described in this paper could be
done only with the help of the single REDUCE system.
  Let us briefly describe the corresponding REDUCE functions
from our package \cite{reduce-package} for applications in meson
chiral models.
   Each REDUCE example below follows the usual mathematical notation.
   \begin{description}
   \item
   [{\tt traceshift(X)}] --
transforms expression to the unique form using the possibility
to perform cyclic permutation of the products of matrix symbols
under the trace operation, e.g.
\begin{eqnarray*}
\tr \left( \partial_\mu\Fi \cdot \Fi \; \partial_\mu\Fi \cdot \Fi \right)
 \, &\longrightarrow& \,
\tr \left( \partial_\mu\Fi \cdot \Fi \; \partial_\mu\Fi \cdot \Fi \right) \\
   \mbox{\tt traceshift(fi(mu)*fi()*fi(mu)*fi())} \, &\longrightarrow& \,
   \mbox{\tt fi(mu)*fi()*fi(mu)*fi()} \\
\tr \left( \Fi \; \partial_\mu\Fi \cdot \Fi \; \partial_\mu\Fi \right)
 \, &\longrightarrow& \,
\tr \left( \partial_\mu\Fi \cdot \Fi \; \partial_\mu\Fi \cdot \Fi \right) \\
   \mbox{\tt traceshift(fi()*fi(mu)*fi()*fi(mu))} \, &\longrightarrow& \,
   \mbox{\tt fi(mu)*fi()*fi(mu)*fi()} \\
\end{eqnarray*}
   \item
   [{\tt orderind(X)}] --
transforms expression to the unique form using the
possibility to redesignate the dummy indices, e.g.
\begin{eqnarray*}
\tr \left( \partial_\mu\Fi \cdot \Fi \; \partial_\mu\Fi \cdot \Fi \right)
 \, &\longrightarrow& \,
\tr \left( \partial_\mu\Fi \cdot \Fi \; \partial_\mu\Fi \cdot \Fi \right) \\
   \mbox{\tt orderind(fi(mu)*fi()*fi(mu)*fi())} \, &\longrightarrow& \,
   \mbox{\tt fi(mu)*fi()*fi(mu)*fi()} \\
\tr \left( \partial_\nu\Fi \cdot \Fi \; \partial_\nu\Fi \cdot \Fi \right)
 \, &\longrightarrow& \,
\tr \left( \partial_\mu\Fi \cdot \Fi \; \partial_\mu\Fi \cdot \Fi \right) \\
   \mbox{\tt orderind(fi(nu)*fi()*fi(nu)*fi())} \, &\longrightarrow& \,
   \mbox{\tt fi(mu)*fi()*fi(mu)*fi()}.
\end{eqnarray*}
   \end{description}

   After voluminous computations one gets the complex expressions
for heat-coefficients $h_1, \ldots h_4$:
\begin{eqnarray*}
h_{0}(x) &=& 1 ,\\
h_{1}(x) &=& -a ,\\
\trp [h_{2}(x)] &=& \trp \left\{
   {1\over 12}(\Gamma_{\mu\nu})^{2}
 + {1\over 2}a^{2}\right\},
\\
\trp [h_{3}(x)] &=& - {1\over 12} \trp \left\{
  {2a^{3}
 - S_\mu S^\mu
 + a(\Gamma_{\mu\nu})^{2}
 - {2\over 45}(K_{\alpha\beta\gamma })^{2}
 - {1\over 9}(K^\alpha {}_{\alpha\beta})^{2}}\right\},
\\
\trp [h_{4}(x)] &=& \trp \Bigg\{
   {1\over 24}a^{4}
 + {1\over 12}{a^{2} S^\mu {}_\mu
 + a S_\mu S^\mu }
 + {1\over 720}{7(S^\mu {}_\mu )^{2}
 -(S_{\mu\nu})^{2}}
\\&&
 + {1\over 30}a^{2}(\Gamma_{\mu\nu})^{2}
 + {1\over 120}(a\Gamma_{\mu\nu})^{2}
 + {1\over 180}a(K^\alpha {}_{\alpha\mu})^{2}
 + {1\over 75}a\Gamma_{\mu\nu}K_\beta {}^{\beta\mu\nu}
 + {7\over 900}\Gamma_{\mu\nu}S^\mu K_\alpha {}^{\alpha\nu}
\\&&
 + {1\over 50}aK_\beta {}^{\beta\mu\nu}\Gamma_{\mu\nu}
 - {1\over 300}\Gamma_{\mu\nu}K_\alpha {}^{\alpha\mu}S^\nu
 + {1\over 3600}K^\alpha {}_{\alpha\mu}
      \left(S_\beta {}^{\beta\mu} + S_\beta{}^{\mu\beta} \right)
 + {1\over 72}S_\mu {}^\mu (\Gamma_{\alpha\beta})^{2}
\\&&
 + {1\over 180}S^{\mu\nu}\{\Gamma_{\mu\alpha},\Gamma_\nu {}^\alpha \}
 + {1\over 40}a \left(\Gamma_{\mu\nu}S^{\mu\nu}
                      + {11\over 9}S_{\mu\nu}\Gamma^{\mu\nu} \right)
 + {1\over 144}a \left[K^\mu {}_{\mu\nu},S^\nu \right]
\\&&
 + \left(  {2\over 135}aK_{\beta\mu\nu}
         + {11\over 900}\Gamma_{\mu\nu}S_\beta
         + {1\over 100}S_\beta \Gamma_{\mu\nu}
         + {1\over 4725}\left[\Gamma_{\mu\nu},K^\alpha_{\alpha\beta}\right]
  \right)
  (K^{\beta\mu\nu} -K^{\mu\nu\beta})
\\&&
 + {1\over 1260} \Gamma_{\mu\nu} K_\alpha {}^{\alpha\mu} K_\beta{}^{\beta\nu}
 - {1\over 12600} \left(
         29\Gamma^\beta {}_\alpha \Gamma^{\mu\alpha}
       + 27\Gamma^{\mu\alpha}\Gamma^\beta {}_\alpha
    \right)
    \left( K^\nu_{\mu\beta\nu}
          + K^\nu_{\mu\nu\beta} \right)
\\&&
 + \Gamma_{\alpha\beta}\Gamma_{\mu\nu}
       \left({83\over 25200}K^{\mu\nu\alpha\beta}
           + {4\over 1575}K^{\alpha\beta\mu\nu}
           - {127\over 5040}K^{\alpha\mu\nu\beta}
           - {1\over 600}K^{\mu\alpha\beta\nu}
       \right)
\\&&
 + {13\over 12600}\Gamma_{\mu\beta} \Gamma^\beta{}_\nu \Gamma^\nu{}_\alpha
 \Gamma^{\alpha\mu}
 + {47\over 16800}(\Gamma_{\mu\nu})^{2}(\Gamma_{\alpha\beta})^{2}
 + {17\over 25200}(\Gamma_{\mu\nu}\Gamma_{\alpha\beta})^{2}
\\&&
 + {4\over 1575}(\Gamma_{\mu\alpha}\Gamma^\alpha{}_\nu)^{2}
 + {19\over 25200}K^\alpha{}_{\alpha\mu\nu}K^\mu{}_\beta{}^{\beta\nu}
 - {1\over 12600}(K^\alpha{}_{\mu\nu\alpha})^{2}
 + {1\over 1575}(K_\mu{}^\alpha{}{}_{\alpha\nu})^{2}
\\&&
 + {1\over 6300} K_\mu{}^\alpha{}_{\alpha\nu} K^{\beta\mu\nu}{}_\beta
 + {1\over 5600}(K^\alpha{}_{\alpha\mu\nu})^{2}
 - {1\over 5040}{K^\alpha{}_{\alpha\mu\nu}K_\beta{}^{\mu\nu\beta}
 + K_{\mu\nu\alpha\beta}K^{\alpha\mu\nu\beta}}
\\&&
 - {1\over 1800} K_{\mu\alpha}{}^\alpha{}_\nu K^{\mu\beta}{}_\beta{}^\nu
 - {1\over 25200} K_{\mu\nu\al\be}
      \left[
       3 \left( K^{\mu\nu\al\be} + K^{\nu\al\be\mu} \right)
     + 2 \left( K^{\mu\al\be\nu} + K^{\al\nu\mu\be} \right)
      \right]
\Bigg\}
\end{eqnarray*}

   In the same way the next orders of heat expansion coefficients
$h_i$ can be obtained using the developed calculation technique
based on the usage of computer algebra.
   For simplicity we present below expressions only for minimal
parts of heat-coefficients, i.e.\ only for the parts which do
not vanish in the pseudoscalar region of the theory when $V_\mu=A_\mu=0$:
\begin{eqnarray*}
\trp [h_{5}(x)^{min}] &=&
 - \trp \Bigg\{
  {1\over 120} a^2 (  a^3
                    + S_\mu S^\mu)
 + {1\over 180} a^3 S_\mu{}^\mu
 + 2(a S_\mu)^2
\\&&
 + {1\over 6300}\left[  10 a S_\mu (S^{\mu\nu}{}_\nu
                       + S_\nu{}^{\nu\mu} )
                       - 2a (S_{\mu\nu})^{2}
                       + 17a (S^\mu_\mu )^{2}
                       + S^{\mu\nu}{}_\nu S_\mu
                       + 3 a S_\mu{}^{\mu\nu} S_\nu
                \right]
\\&&
 + {11\over 1008}S_\mu S^\mu S_\nu{}^\nu
 + {19\over 2800}S_\mu S_\nu S^{\mu\nu}
 + {2\over 225}S_\mu S_\nu S^{\nu\mu}
\\&&
 + {1\over 25200}\left[ {3(S^\mu{}_{\mu\nu})^{2}
 -2(S_{\mu\nu\alpha})^{2}
 -23(S^{\mu\nu}{}_\nu )^{2}
 + 7S^\mu{}_{\mu\nu} S^{\nu\alpha}{}_\alpha }\right] \Bigg\}
\\
\trp [h_{6}(x)^{min }] &=& \trp \Bigg\{ {1\over 720}a^{2}(a^{4}
 + 4S_\mu a S^\mu )
 + {1\over 420}a^{3}S_\mu S^\mu
\\&&
 + {1\over 20160}a^{2}\left[
   20a^{2}S_\mu{}^\mu
 + 5S_\mu (S^{\mu\nu}{}_\nu
 + S_\nu{}^{\nu\mu} )
 + S^\mu{}_{\mu\nu}S^\nu
 -(S_{\mu\nu})^{2}
 + 11(S_\mu{}^\mu )^{2}
 + 9S^{\mu\nu}{}_\nu S_\mu
                       \right]
\\&&
 + {1\over 25200}a\left[{S_\mu a(73S^{\mu\nu}{}_\nu
 + 37S_\nu{}^{\nu\mu} )
 + 5S^{\mu\nu}(S_\mu S_\nu
 + 4S_\nu S_\mu )}\right]
 + {1\over 2016} a S^\mu{}_\mu S^\nu S_\nu
\\&&
 + {1\over 9450} a S_\mu \left(  37S^{\mu\nu}
                               + 23S^{\nu\mu}
                         \right) S_\nu
 + {1\over 9072} a S_\mu
        \left(  S^{\mu\nu}{}_\nu{}^\alpha{}_\alpha
              + S_\nu{}^{\nu\mu\alpha}{}_\alpha
              + S_\nu{}^\nu{}_\alpha{}^{\alpha\mu}
        \right)
\\&&
 + a S^\mu \left(  {23\over 4800} S_\mu S_\nu{}^\nu
                 + {937\over 302400} S^\nu S_{\mu\nu}
                 + {23\over 10800} S^\nu S_{\nu\mu}
            \right)
 + {1\over 252} a S_\mu S^\nu{}_\nu S^\mu
\\&&
 + {1\over 352800}aS^\mu{}_{\mu\nu} \left(  52S_\alpha{}^{\alpha\nu}
                                          + 53S^{\nu\alpha}{}_\alpha
                                    \right)
 + aS^{\mu\nu}{}_\nu \left(  {1\over 3360}S^\alpha{}_{\alpha\mu}
                            -{1\over 11340}S_{\mu\alpha}{}^\alpha
		     \right)
\\&&
 + {17\over 226800}a(S_{\mu\nu\alpha})^{2}
 + {1\over 317520}a \left(  23S^\mu{}_\mu{}^{\nu\alpha}{}_\alpha
                          + 5S^\mu{}_{\mu\alpha}{}^{\alpha\nu}
                          + 77S^{\nu\mu}{}_\mu{}^\alpha{}_\alpha
		    \right) S_\nu
\\&&
 - {1\over 30240}\left[{53(S_\mu S^\mu )^{2}
 + (S_\mu S_\nu )^{2}}\right]
 + {1\over 352800}S_\mu S_\nu \left(  157S^\mu{}_\alpha{}^{\al\nu}
                                    + 298S^{\mu\nu\alpha}{}_\alpha
			      \right)
\\&&
 + {1\over 70560}S_\mu S_\nu \left(  31S_\alpha{}^{\alpha\nu\mu}
				   + 58S^{\nu\mu\alpha}{}_\alpha
				   + 47S^{\nu\alpha}{}_\alpha{}^\mu
			     \right)
 + {1\over 720}S^\mu S_\mu S^\nu{}_\nu{}^\alpha{}_\alpha
\\&&
 + {5\over 14112}S_\mu S_\nu S_\alpha{}^{\alpha\mu\nu}
 + {1\over 105840}S_\mu \left[  {S^{\mu\nu} \left(  37S^\alpha{}_{\alpha\nu}
                                                  + 70S_{\nu\alpha}{}^\alpha
					    \right)
                              + 35S^{\nu\mu}S^\alpha_{\nu\alpha}}
			\right]
\\&&
 + {1\over 21168}S_\mu \left[  {S^{\nu\mu}S^\alpha{}_{\alpha\nu}
			     + S_{\nu\alpha}(5S^{\nu\mu\alpha}
			     + 2S^{\nu\alpha\mu})}
 			\right]
 + {1\over 2880}S_\mu \left(  S^{\mu\nu}{}_\nu
                            + S_\nu{}^{\nu\mu}
			\right) S_\alpha{}^\alpha
\\&&
 + S_\mu S^\nu{}_\nu \left(  {83\over 141120}S^{\mu\alpha}{}_\alpha
                           + {1\over 9408}S_\alpha{}^{\alpha\mu}
		     \right)
 + {1\over 30240}S_\mu \left(  17S_{\nu\alpha}{}^\alpha
                             + 13S^\alpha{}_{\alpha\nu}
			\right) S^{\nu\mu}
\\&&
 + {1\over 7560}S_\mu \left[  2 \left(  S^{\mu\nu\alpha}
			              + S^{\nu\mu\alpha}
			              + S^{\nu\alpha\mu}
                                \right) S_{\nu\alpha}
			    + \left( S_{\nu\alpha}{}^\alpha
			            + 2S^\alpha{}_{\alpha\nu}
                              \right) S^{\mu\nu}
		       \right]
 + {1\over 2160}S_\mu S_{\nu\alpha}S^{\mu\nu\alpha}
\\&&
 - {1\over 635040} \left(  701(S_\mu{}^\mu )^{3}
                         + 583S^{\mu\nu}S_{\mu\alpha}S^\alpha_\nu
		   \right)
 - {689\over 316386}S^\alpha{}_\alpha (S^{\mu\nu})^{2}
\\&&
 - {2\over 2835}S^\mu{}_\mu S^{\nu\alpha} S_{\alpha\nu}
 - {1\over 952560}S^{\mu\nu} \left(  619 S^\alpha{}_\nu S_{\mu\alpha}
                                   + 190 S_{\mu\alpha} S^\alpha{}_\nu
			     \right)
\\&&
 + {1\over 151200} \left[  11 (S^\mu{}_\mu{}^\nu{}_\nu)^2
                         - 2 (S_{\mu\nu}{}^\alpha{}_\alpha)^2
                   \right]
 + {1\over 176400} \left(  (S^\mu{}_{\mu\nu\alpha})^{2}
                         + S^\mu{}_{\mu\nu\alpha} S^{\nu\beta}{}_\beta{}^\alpha
		   \right)
\\&&
 - {1\over 226800}(S_{\mu\nu\alpha\beta})^{2}
 - {103\over 12700800}(S_\mu{}^\alpha{}_{\alpha\nu})^{2}
+ {1\over 66150}S^\mu{}_{\mu\nu\alpha} S^{\nu\alpha\beta}{}_\beta
\\&&
 - {1\over 52920} S^\mu{}_\mu{}^\nu{}_\nu S^{\alpha\beta}{}_{\beta\alpha}
 - {13\over 604800} S^{\mu\nu\alpha}{}_\alpha  S_{\mu\beta}{}^\beta{}_\nu
      \Bigg\} .
\end{eqnarray*}

   To obtain these expressions for the heat coefficients we have
extensively made use of the cyclic properties of the trace and
of the Jacobi identity.
   To obtain effective meson Lagrangians in terms of collective
fields one should calculate in $\trp h_i(x)$ the trace over
Dirac indices.
\\
\\
   We would like to thank V.P.Gerdt for useful discussions.
Two of the us (A.A.Bel'kov and A.V.Lanyov) are grateful for the
hospitality extended to them at DESY-Zeuthen.

\begin{thebibliography}{99}
   %
   \bibitem{form} J.A.M.Vermaseren, FORM Manual, version 1, 1991
   %
   \bibitem{reduce} A.C.Hearn, REDUCE User's Manual, version 3.4,
                    RAND Publication CP78 (Rev.7/91), Santa Monica, July 1991
   %
   \bibitem{91-08} D.Ebert, A.A.Bel'kov, A.V.Lanyov and A.Schaale,
 Preprint PHE 91-08, Zeuthen, 1991
   %
   %
   \bibitem{qcd} H.Fritzsch, M.Gell-Mann, H.Leutwyler, Phys. Lett. B47 (1973)
365
   %
   \bibitem{njl} Y.Nambu, G.Jona-Lasinio, Phys. Rev. 122 (1961) 345
   %
   \bibitem{ebert-reinhardt} D.Ebert and H.Reinhardt, Nucl. Phys. B271 (1986)
188
   %
   \bibitem{kikkawa} K.Kikkawa, Progr. Theor. Phys. 56 (1976) 947
   %
   \bibitem{volkov-ebert}
                                M.K.Volkov, Yad. Fiz. 6 (1967) 1100;
                                          7 (1968) 445;
                                Ann. of Phys. (N.Y.) 49 (1968) 202;
                                Fortschr. Phys. 28 (1974) 499
   %
   \bibitem{reduce-package} A.A.Bel'kov and A.V.Lanyov, JINR
 Communication E11-91-162, Dubna, 1991;\\
                            Proc. ISSAC'91, July 1991, Bonn,
 Germany, ed. S.M.Watt, ACM Press, 1991, p.454
   %
   \end{thebibliography}
   \end{document}